\documentclass[12pt]{article}
\usepackage{amssymb}
\usepackage{amsmath}
\begin{document}
\begin{center}
{\bf CASIMIR FRICTION FORCE AND ENERGY DISSIPATION FOR MOVING HARMONIC OSCILLATORS. II}

\vspace{1cm}
Johan S. H{\o}ye\footnote{johan.hoye@ntnu.no}

\bigskip
Department of Physics, Norwegian University of Science and Technology, N-7491 Trondheim, Norway

\bigskip
Iver Brevik\footnote{iver.h.brevik@ntnu.no}

\bigskip
Department of Energy and Process Engineering, Norwegian University of Science and Technology, N-7491 Trondheim, Norway

\bigskip
\end{center}

\begin{abstract}

This paper is a second in a series devoted to the study of a two-oscillator system in linear relative motion
  (the first one published as a letter in Europhys. Lett. {\bf 91}, 60003 (2010)). The main idea behind considering this kind of system is to use it as a simple model for Casimir friction. In the present paper we extend our previous theory so as to obtain the change in the oscillator energy to  second order in the perturbation, even though we employ first order perturbation theory only. The results agree with, and confirm, our earlier results obtained via different routes. The friction force is finite at finite temperatures, whereas in
  the case of two oscillators moving with constant relative velocity the force becomes zero at zero temperature, due to slowly varying coupling.

\end{abstract}


\section{Introduction}

Consider two dielectric or metallic slabs with parallel surfaces,
closely separated. If the slabs are set in tangential motion with
respect to each other, with constant velocity, they become exposed
to a friction force called Casimir friction, an effect that has
received considerable attention in the recent past. The physical
reason for the effect is that photons transferred between the
slabs are subject to Doppler shifts. Such frequency shifts are
physically  expected to lead to energy dissipation, and hence a
friction force. Some papers, limited
 to the period from 2007 onwards, are listed in refs.~\cite{pendry97,pendry98,volokitin07,volokitin08,philbin09,philbin09a,dedkov08,dedkov08a,dedkov10,pendry10}. As one might
  expect, the problem becomes  singular if the relative velocity of the slabs is simply taken to be  constant, from $t=-\infty$ to $t=+\infty$. It is physically advantageous, therefore,
   to  imagine that the interaction is effectively coupled in for a finite period of time.
   We shall make use of this technique in the following. It is in principle similar to the  technique
  often employed in scattering field theory.

 Now it might appear most natural to attack the Casimir friction problem by making use of standard macroscopic electrodynamics for media in constant rectilinear motion. It means
 use of Maxwell's equations in moving matter endowed with a refractive index $n$. And as expected, it turns out that most of the mentioned papers are following this kind of approach.
  A complicating factor in the present case is, however, that there is  no obvious rest system of the matter to refer to;  none of the slabs plays a privileged role. This
  contrasts the usual case in phenomenological electrodynamics where the covariant formulation is simply constructed such that the theory reduces to conventional electrodynamics
   in the uniquely defined rest inertial system (cf., for instance, Refs.~\cite{jauch48,moller72,brevik79}).

   Faced with this circumstance or difficulty, it lies at hand to search for
   alternative approaches to the friction problem. On natural
   possibility is then to make use of statistical mechanical methods,
   for harmonic oscillators in uniform relative motion. These methods were used by us in a recent work \cite{hoye10}, and were used also in previous works \cite{hoye92,hoye93}.  The
   microscopic method has some advantages in comparison with
   macroscopic electrodynamics, most notably that the formalism
   becomes more simple and transparent. Yet, the statistical
   mechanical approach has the property that it is capable of
   showing the main features of the problem. The microscopic approach
   has also been followed recently by Barton (preprint, personal
   communication). In the present paper we follow basically our earlier avenue of research.

   \vspace{0.3cm}

To begin with, let us briefly summarize the basic characteristics of the model we consider (cf. also Ref.~\cite{hoye10}):

The two-oscillator system is quantum-mechanical in nature; the reference state  corresponding to a Hamiltonian $H_0$ with no coupling. Thermal equilibrium is initially assumed. The equilibrium state becomes disturbed by a time-dependent term written as $-Aq(t)$, where $A$ is a time-independent operator and $q(t)$ a classical function of time. Its explicit form depends on the details of the system. The Hamiltonian now becomes $H=H_0-Aq(t)$. As a specific case we considered in Ref.~\cite{hoye10} the situation with two oscillators in relative motion with
\begin{equation}
-Aq(t)=\psi({\bf r})x_1x_2, \label{1A}
\end{equation}
$\psi({\rm r})$ being the coupling strength, $\bf r$  the separation between the oscillators, and $x_1,x_2$ the internal vibrational coordinates. If $\bf v$ denotes the non-relativistic constant relative velocity, the interaction varies as
\begin{equation}
-Aq(t)=[\psi({\bf r}_0)+ {\boldsymbol \nabla}\psi({\bf r}_0)\cdot {\bf v}t+...]x_1x_2, \label{2A}
\end{equation}
when expanded around the initial position ${\bf r}={\bf r}_0$ at $t=0$. The force $\bf B$ between the oscillators is
\begin{equation}
{\bf B}=-({\boldsymbol \nabla}\psi)x_1 x_2. \label{3A}
\end{equation}
However, as mentioned above the situation becomes singular, so to obtain the dissipation the perturbing interaction should last for only a finite period of time. Thus ${\bf v}t$ in Eq.~(\ref{2A}) was replaced by ${\bf v}te^{-\eta t}$ for $t>0 $ (and $0$ for $t<0$) and the limit $\eta \rightarrow 0$ was considered. For that situation we found no dissipation at $T=0$. Then a more general interaction $-Aq(t)$ was considered in Ref.~\cite{hoye10}, and an expression for the dissipation was found where friction could be present also at $T=0$. To arrive at this result we also associated the general situation with motion. In view of the attention paid to the friction problem it is of interest to check the results of Ref.~\cite{hoye10} in an independent way, as we do in the present work.

\vspace{0.3cm}
After this brief survey let us now point out  the chief results obtained in the present paper:

   \begin{itemize}
   \item  We calculate a general expression for the change in energy, called   $\Delta E$,     by means
   of quantum mechanical perturbation theory. This energy change occurs to second order in the perturbation. Nevertheless, time-dependent perturbation theory to the {\it first} order is sufficient to find the second order effect. This is because the phases of the perturbed change in amplitude, and the initial amplitude of each eigenstate, are uncorrelated at thermal equilibrium. Change in the amplitudes of eigenstates will accordingly be the square of perturbed amplitudes; i.e. there are no cross-terms.
   We moreover find that $\Delta E$  is always positive or zero,  corresponding to a
   friction force. Doubts occasionally raised in the literature
   about the very existence of the Casimir friction effect
   \cite{philbin09} are thus from this standpoint laid at rest.

   \item Making use of the mentioned expression for $\Delta E$ we compare the present formalism with that of Ref.~\cite{hoye10}, the latter result obtained in a quite different
   way and on a quite different form. In Ref.~\cite{hoye10}, the linear response via the Kubo formalism was used \cite{hoye92,hoye93,kubo58} to calculate the force which in turn could be divided into a reversible and an irreversible part.  It is the latter part that is associated with dissipation.
    A satisfactory feature
   is that the present derivation, although  being  quite different
 from that of Ref.~\cite{hoye10}, leads to the same physical result.
\end{itemize}

\section{Time-dependent perturbation theory}

To fix the notation, we consider the  perturbation theory for a
system at thermal equilibrium. The wave function can be written as
\begin{equation}
\psi=\sum_na_n\psi_n, \label{1}
\end{equation}
where $\psi_n=\psi_n(x)$ are the eigenstates. For simplicity we here let $x$ represent all the coordinates of the system. If $\psi$ is normalized, $\int \psi^*\psi dx=1$, then $|a_n|^2$ is the probability for the system to be in eigenstate $n$. At thermal equilibrium this probability is given by the Boltzmann factor
\begin{equation}
P_n=|a_n|^2=\frac{1}{Z}\,e^{-\beta E_n}, \label{2}
\end{equation}
where $E_n$ is the energy eigenvalue of the state and $Z$ is the partition function
\begin{equation}
Z=\sum_ne^{-\beta E_n}. \label{3}
\end{equation}
Let now the Hamiltonian be perturbed  by the time-dependent
interaction $V(t)=-Aq(t)$, as mentioned above. Due to the
perturbation the coefficients $a_n$ will change. If the system
starts in a state $m$ there are transitions to other states given
by a change in $a_n$
\begin{equation}
\Delta a_n=b_{nm}. \label{5}
\end{equation}
The $b_{nm}$ is given by the standard expression
\begin{equation}
b_{nm}=\frac{1}{i\hbar}\int_{-\infty}^tV_{nm}(\tau)e^{i\omega_{nm}\tau} d\tau, \label{6}
\end{equation}
where \[ V_{nm}(\tau)=\int \psi_n^*\,
V(\tau)\psi_mdx=-A_{nm}\,q(\tau),
\]
\begin{equation}
A_{nm}=\langle n|A|m\rangle=\int \psi_n^*\,A\,\psi_mdx. \label{7}
\end{equation}
Here $\omega_{nm}=\omega_n-\omega_m$, with $\omega_n=E_n/\hbar$.

As mentioned above, to avoid singularities caused by too idealized
conditions we will assume that the perturbation vanishes after
some time. Then we will obtain the total change in $\Delta a_n$
with
\[
b_{nm}=-\frac{1}{i\hbar}A_{nm}\,\hat{q}(-\omega_{nm}), \]
\begin{equation}
\hat{q}(\omega)=\int_{-\infty}^\infty q(t)e^{-i\omega t}dt,
\label{8}
\end{equation}
where the hat denotes Fourier transform.

From a general perspective, the system may start in a combination of eigenstates with transitions from several states. (It might be natural in this context
 to think about the Casimir-Polder setup with molecules traveling close to a dielectric surface. For molecules, in contrast to atoms, the energy levels are
  closely separated and may thus easily allow transitions. For recent investigations along these lines, cf. Refs.~\cite{buhmann08,ellingsen09}.) With this,
   Eq.~(\ref{5}) will be modified to $\Delta a_n=\sum_{m \neq n}a_mb_{nm}$. Now, the state $n$ does not only receive contributions, but gives away contributions
    to other states also. The latter must follow  from the corresponding increase of probabilities for the other states. Omitting the latter for the moment, the perturbed coefficients
    are
\begin{equation}
a_{1n}=a_n+\Delta a_n=a_n+\sum_{m\neq n}a_mb_{nm}. \label{9}
\end{equation}
The $a_n$ will have complex phase factors, and in thermal
equilibrium one must assume the phases of $a_n$ and $a_m$ $(m\neq
n$) to be uncorrelated. Thus by thermal average
\begin{equation}
\langle a_n^* a_m\rangle =0. \label{110}
\end{equation}
With this the new probability of the state $n$ becomes
\[
P_{1n}=\langle a_{1n}^* a_{1n}\rangle=|a_n|^2+\sum_{m\neq n}|a_m|^2B_{nm}, \]
\begin{equation}
B_{nm}=b_{nm}b_{nm}^* =|b_{nm}|^2. \label{11}
\end{equation}
The last term is the increase in probability from the other states. Likewise, the state $n$ must obey a similar loss of probability to other states to conserve probability.
 The loss to other states is thus $\sum_{m\neq n}|a_n|^2B_{mn}$. With Eq.~(\ref{6}) we have $b_{mn}=b_{nm}^*$, by which $B_{mn}=B_{nm}$. The latter equation reflects that
 the transition probabilities between each pair of states are the
 same in either direction. With this, the resulting perturbed
 probability of state $n$ modifies the expression (\ref{11}) into
 \begin{equation}
  P_{1n}=|a_n|^2+\sum_{m\neq
 n}(|a_m|^2-|a_n|^2)B_{nm}=P_n+\sum_m(P_m-P_n)B_{nm}. \label{12}
 \end{equation}
 The change in energy can now be evaluated as
 \[ \Delta E=\sum_nE_n(P_{1n}-P_n)=\sum_{nm}E_n(P_m-P_n)B_{nm} \]
 \begin{equation}
 =\sum_{nm}(E_n-E_m)P_m B_{nm}+\sum_{nm}(E_m P_m-E_n P_n)B_{nm}=\sum_{nm}(E_n-E_m)P_m B_{nm}.
 \label{13}
 \end{equation}
Utilizing the symmetry with respect to $n$ and $m$ in this
expression and inserting for $P_m$ from Eq.~(\ref{2}) we find
\begin{equation}
\Delta
E=\frac{1}{2}\sum_{nm}(E_n-E_m)(P_m-P_n)B_{nm}=\frac{1}{Z}\sum_{nm}e^{-\frac{1}{2}\beta
(E_n+E_m)}\Delta_{nm}\sinh (\frac{1}{2}\beta \Delta_{nm})B_{nm},
\label{14}
\end{equation}
with $\Delta_{nm}=E_n-E_m$, and where from Eqs.~(\ref{8}) and
(\ref{11})
\begin{equation}
B_{nm}=\frac{1}{\hbar^2}A_{nm}A_{nm}^*\hat{q}(-\omega_{nm})\hat{q}(\omega_{nm}).
\label{15}
\end{equation}
Here it is to be noted that $\Delta E \geq 0$. We conclude that
whenever a system in thermal equilibrium is disturbed by some
external perturbation the energy always increases (or, is
unchanged), i.e., energy is dissipated. The dissipation occurs to
second order in the perturbation. To first order there is no
dissipation; the changes are adiabatic.

We note in passing the similar nature of the formalism for an
electromagnetic field in a dissipative medium: the mean quantity
of heat developed per unit time and volume is
\[ Q=\omega \varepsilon''(\omega)\langle {\bf E}^2 \rangle, \]
where $\varepsilon''$ denotes the imaginary part of the
permittivity, $\varepsilon=\varepsilon'+i\varepsilon''$.
Irreversibility of the dissipation process implies the condition
$\varepsilon''(\omega)>0$ for positive $\omega$. Cf., for
instance, ref.~\cite{landau84}, Sec. 80.

\section{Energy dissipation from friction force}

In Ref.~\cite{hoye10}, we evaluated the dissipated energy by considering the
friction force. The $q(t)$ denotes (or can be
interpreted to denote) the position
\begin{equation}
x=x(t)=q(t), \label{16}
\end{equation}
and  $A$ becomes correspondingly the operator for a force. By use
of the Kubo relation \cite{hoye92,hoye93,kubo58} for this
situation the resulting force due to the perturbation is
\begin{equation}
F_f=\int_{-\infty}^t \phi_{AA}(t-t')q(t')dt', \label{17}
\end{equation}
where
\begin{equation}
\phi_{AA}(t)=\frac{1}{i\hbar}\rm{Tr}\left\{ \rho [A, A(t)]\right\}.\label{18}
\end{equation}
Here
\[
\rho=\frac{e^{-\beta H}}{Z}, \quad \rm{with} \quad  Z=\rm{Tr}(e^{-\beta H}), \]
is the canonical density matrix, and
\begin{equation}
A(t)=e^{itH/\hbar}A\,e^{-itH/\hbar}, \label{19}
\end{equation}
with $H$ the Hamiltonian. With velocity
\begin{equation}
v(t)=\dot{x}(t)=\dot{q}(t) \label{20}
\end{equation}
the total energy dissipated by the system is
\begin{equation}
\Delta E=-\int_{-\infty}^\infty v(t)F_f dt=-\int_{-\infty} ^\infty \left[ \int_{-\infty}^t \dot{q}(t)\phi_{AA}(t-t')q(t')dt' \right] dt, \label{21}
\end{equation}
which is the same result as in Eq.~(27) in Ref.~\cite{hoye10}. Now the quantity $q(t)$ need not be a position as given by Eq.~(\ref{16}), but as it can be interpreted as
 a position in a generalized sense, we concluded in \cite{hoye10} that the result (\ref{21}) has a broader applicability. We will now show that this is
  actually the case, by showing that the result (\ref{21}) is the
  same as the new result (\ref{14}) obtained in the present work, by
  means of time-dependent
perturbation theory.

 With wave function representation we first have
 \begin{equation}
 e^{-\beta H} \rightarrow \sum_n\psi_n(x)e^{-\beta
 E_n}\psi_n^*(x_1), \label{22}
 \end{equation}
 \[
 \rho A A(t)=\frac{1}{Z}\sum_ {nmk}\int \psi_n(x)e^{-\beta
 E_n}\psi_n^*(x_1)A\psi_m(x_1)e^{i\omega_mt}\psi_m^*(x_2)A \]
 \begin{equation}
 \times \psi_k(x_2)e^{-i\omega_kt}\psi_k^*(x_3)dx_1dx_2.
 \label{23}
 \end{equation}
 Thus we obtain
 \begin{equation}
 {\rm Tr}(\rho AA(t))=\frac{1}{Z}\sum_{nm}e^{-\beta
 E_n}A_{nm}e^{i\omega_mt}A_{mn}e^{-i\omega_nt}, \label{24}
 \end{equation}
 as $\int \psi_k^*(x)\psi_n(x)dx=\delta _{kn}$ $ (x_3=x_1=x)$, and
 $A_{nm}$ is given by Eq.~(\ref{7}). Likewise we calculate ${\rm
 Tr}(\rho A(t)A)$ by exchange of $\omega_n$ and $\omega_m$ in
 Eq.~(\ref{24}). The response function becomes
 \begin{equation}
 \phi_{AA}(t)=\frac{1}{i\hbar}{\rm Tr}\left\{ \rho [A, A(t)]\right
 \}=\frac{1}{i\hbar}\sum_{nm}M_{nm}(e^{-i\omega_{nm}t}-e^{i\omega_{nm}t}),
 \label{25}
 \end{equation}
 with
 \begin{equation}
 M_{nm}=-\frac{1}{Z}e^{-\frac{1}{2}\beta (E_n+E_m)}\sinh (\frac{1}{2}\beta
 \Delta_{nm})A_{nm}A_{nm}^* \label{26}
 \end{equation}
 (recall that $\Delta_{nm}=E_n-E_m=\hbar \omega_{nm},
 ~A_{mn}=A_{nm}^*)$.  The expression for $M_{nm}$ follows
 if  one first exchanges  $n$ and $m$ in Eq.~(\ref{24}), then adds the
 resulting term to it and divides by 2. By inserting
 Eq.~(\ref{25}) into Eq.~(\ref{21}) one gets the integral
 \begin{equation}
 I=\int_{t>t'} \int \dot{q}(t)q(t')\left( e^{-i\omega
 t}e^{i\omega t'}-e^{i\omega t}e^{-i\omega t'}\right)dt'dt,
 \label{27}
 \end{equation}
 where here
 $\omega=\omega_{nm}~[(E_n-E_m)/\hbar=\Delta_{nm}/\hbar]$. By
 partial integration and exchange of integration variables $t$ and
 $t'$ (in the last term below ) we get
 \[ I=i\omega \int_{t>t'}\int q(t)q(t')\left( e^{-i\omega
 t}e^{i\omega t'}+e^{i\omega t}e^{-i\omega t'}\right) dt'dt \]
 \begin{equation}
 =i\omega \int_{-\infty}^\infty \int_{-\infty}^\infty
 q(t)q(t')e^{-i\omega t}e^{i\omega t'}dt'dt=i\omega
 \hat{q}(\omega)\hat{q}(-\omega). \label{28}
 \end{equation}
 By inserting this into Eq.~(\ref{21}) via Eq.~(\ref{25}) we get
 for the dissipated energy
 \begin{equation}
 \Delta E=\frac{1}{\hbar}\sum_{nm}M_{nm}\,\omega \,
 \hat{q}\,(\omega)\,\hat{q}(-\omega). \label{29}
 \end{equation}
 With $\omega=\omega_{nm}=\Delta_{nm}/\hbar$ and $M_{nm}$ given by
 the expression (\ref{26}) this is nothing but the result
 (\ref{14}) together with (\ref{15}) obtained by time-dependent
 perturbation theory. Thus we have been able to derive the same expression for the dissipated energy in two independent ways.

\section{Friction between harmonic oscillators}

In Ref.~\cite{hoye10} the friction between a pair of harmonic
oscillators with interaction as in Eq.~(\ref{2A}) $(t>0)$ was
evaluated. Here, we will evaluate the energy dissipation by a
direct use of Eq.~(\ref{14}) or (\ref{29}). The first term in
Eq.~(\ref{2A}) can be disregarded as it gives a reversible force,
distinct from dissipation. Further, we replace $t$ with $te^{-\eta
t}$ ($\eta \rightarrow 0$) to make the interaction vanish as
$t\rightarrow \infty$.

For harmonic oscillators one can introduce the usual annihilation and creation operators
\begin{equation}
x_i=\left( \frac{\hbar}{2m_i\omega_i}\right)^{1/2}(a_i+a_i^\dagger) \label{31}
\end{equation}
$(i=1,2)$, with properties
\[ a^\dagger |n\rangle=\sqrt{n+1}\,|n+1\rangle, \]
\begin{equation}
a|n\rangle=\sqrt{n}\,|n-1\rangle. \label{32}
\end{equation}
With this the interaction becomes
\begin{equation}
-Aq(t)=\gamma (a_1a_2+a_1a_2^\dagger +a_1^\dagger a_2+a_1^\dagger a_2^\dagger)\,te^{-\eta t}, \label{33}
\end{equation}
where
\begin{equation}
\gamma=(\frac{1}{2}D\hbar)^{1/2}({\bf v\cdot \nabla}\psi), \quad D=\frac{\hbar}{2m_1m_2\omega_1\omega_2}. \label{34}
\end{equation}
Since here only small $\eta$ ($\rightarrow 0$) is considered, the terms $a_1a_2$ and $a_1^\dagger a_2^\dagger$ will not contribute. Thus we can use
\begin{equation}
A=a_1a_2^\dagger +a_1^\dagger a_2, \quad {\rm and} \quad q(t)=\gamma te^{-\eta t}. \label{35}
\end{equation}
For the matrix elements (\ref{7}) we then get
\[ A_{n_1,n_2,n_1+1,n_2-1}=\langle n_1n_2|a_1a_2^\dagger|n_1+1,n_2-1\rangle =\sqrt{n_1+1}\,\sqrt{n_2}, \]
\begin{equation}
A_{n_1,n_2,n_1-1,n_2+1}=\langle n_1n_2|a_1^\dagger a_2|n_1-1,n_2+1\rangle=\sqrt{n_1}\, \sqrt{n_2+1}, \label{36}
\end{equation}
while all other elements are zero. The Fourier transform of $q(t)$ is, for $t>0$,
\begin{equation}
\hat{q}(\omega)=\gamma \int_0^\infty te^{-\eta t}e^{-i\omega t}=\frac{\gamma}{(\eta+i\omega)^2}, \label{37}
\end{equation}
so that, for $\eta \rightarrow 0$,
\begin{equation}
\hat{q}(\omega)\hat{q}(-\omega)=\frac{\gamma^2}{(\eta^2+\omega^2)^2}
 \rightarrow  \frac{\pi \gamma^2}{2\eta \omega^2}\delta(\omega). \label{38}
\end{equation}
here $\omega=\omega_1-\omega_2$, where $\omega_1$ and $\omega_2$ are the eigenfrequencies of the two oscillators. Further, with $\omega \rightarrow 0$ $(m=n \pm 1)$
\begin{equation}
\Delta_{nm}\sinh (\frac{1}{2}\beta \Delta_{nm})\rightarrow \frac{1}{2}\beta \Delta_{nm}^2=\frac{1}{2}\beta(\pm \hbar \omega)^2=\frac{1}{2}\beta \hbar^2\omega^2. \label{39}
\end{equation}
Then the matrix elements (\ref{36}) should be squared and averaged
by the Boltzmann distribution given by Eq.~(\ref{14}). We have
$\langle n_1\rangle \approx \langle n_2\rangle \approx \langle
n\rangle,$ with $\omega_1 \rightarrow \omega_2$, and
\[ \langle n\rangle=\frac{\sqrt{x}}{Z}\sum_{n=0}^\infty nx^n=\frac{x}{1-x}, \quad x=e^{-\beta \hbar \omega_1}, \]
\begin{equation}
Z=\sqrt{x}\sum_{n=0}^\infty x^n=\frac{\sqrt{x}}{1-x}. \label{40}
\end{equation}
Then $\langle n\rangle +1=1/(1-x)$, by which
\[
\langle (n_1+1)n_2+n_1(n_2+1)\rangle=2(\langle n\rangle+1)\langle n\rangle \]
\begin{equation}
=\frac{2x}{(1-x)^2}=\frac{1}{2\sinh^2(\frac{1}{2}\beta \hbar \omega_1)}. \label{41}
\end{equation}
Inserting in Eqs.~(\ref{14}) and (\ref{15}) by multiplying together Eqs.~(\ref{38}), (\ref{39}), and (\ref{41}) we obtain for the energy dissipation
\begin{equation}
\Delta E=\frac{\pi \beta \hbar^2\gamma^2}{8\eta \sinh^2(\frac{1}{2}\beta \omega_1)}\, \delta (\omega_1-\omega_2). \label{42}
\end{equation}
With $\gamma$ inserted from Eq.~(\ref{34}) this is the same as the result (21) of Ref.~\cite{hoye10} with its Eq.~(19) for the friction force inserted.

According to Eq.~(\ref{42}), the case of zero temperature ($\beta
\rightarrow \infty$) yields $\Delta E \rightarrow 0$. This result
is related to our assumptions, including slowly varying coupling
or low velocities, i.e. $\eta \rightarrow 0$ in Eqs.~(\ref{35})
and (\ref{38}). At more rapidly varying coupling or higher
 velocities also finite frequencies
would contribute, leading to a finite energy change and a finite
friction force at $T=0$.

\section{Summary}

Let us summarize our work as follows (cf. also the end of Sect. 1):
\vspace{0.3cm}

{\bf 1.}  The total energy dissipation was calculated for a system perturbed by a time-dependent interaction. The change in energy is basically a second order effect but it was calculated with the use of standard time-dependent perturbation theory only, the reason being the absence of cross-terms due to uncorrelated phases of eigenstates. The energy change was always found to be positive or zero. The result agrees with, and confirms, our previous result of Ref.~\cite{hoye10} obtained in a different and independent way.

{\bf 2.} Our results are basically worked out at any temperature, and we have assumed initial thermal equilibrium. Moreover, we have assumed low velocities and nonrelativistic mechanics. Photons are accordingly not present in the theory.
Photons were included, however, in our earlier study
\cite{hoye93}.

{\bf 3.} The energy change $\Delta E$ is finite in general. This corresponds to a finite friction force. In the limit $T\rightarrow 0$ the expression (\ref{42}) gives, however, $\Delta E \rightarrow 0$. This result is due to our  assumption about constant velocity, involving slowly varying coupling. For couplings  varying more rapidly, there will also be a friction force at $T=0$, due to transitions to excited states.


\end{document}